\documentclass[sn-aps]{sn-jnl}% American Physical Society (APS) Reference Style

\usepackage{braket}
\usepackage{cases}
\usepackage{bm}
\usepackage{wrapfig}
\usepackage{subcaption}
\usepackage{comment}
\usepackage{mathtools}
\jyear{2021}%

\theoremstyle{thmstyleone}%
\theoremstyle{thmstyletwo}%

\theoremstyle{thmstylethree}%

\begin{document}

\title[Analysis of shape change of droplet in dipolar Bose-Hubbard model]{Analysis of shape change of droplet in dipolar Bose-Hubbard model}
%%GivenName	-> \fnm{Joergen W.}

\author*[]{\fnm{Kazuhiro} \sur{Tamura}}%\email{iauthor@gmail.com}

\author[]{\fnm{Shohei} \sur{Watabe}}%\email{iiauthor@gmail.com}

\author[]{\fnm{Tetsuro} \sur{Nikuni}}%\email{iiiauthor@gmail.com}

\affil[]{\orgdiv{Department of physics}, \orgname{Tokyo University of Science}}
%\affil*[1]{\orgdiv{Department of physics}, \orgname{Tokyo University of Science}, \orgaddress{\street{Street}, \city{City}, \postcode{100190}, \state{State}, \country{Country}}}


\abstract{The long-range and anisotropic nature of the dipolar interaction provides the so-called supersolid phases in Bose-Einstein condensates (BECs) in an optical lattice. However, in a certain area of dipole interaction parameters, BECs can form into a droplet. In this paper, in order to qualitatively understand the droplet formations, we propose a toy model that allows us to estimate the size and shape of droplets in dipolar Bose-Hubbard system in the optical lattice. We compare results of the toy model with numerical solutions of the mean-field calculation.}

\maketitle

\section{Introduction}\label{sec:Introduction}
Ultracold Bose gases have been studied to establish fundamental physics of various quantum phenomena due to their flexible degrees of freedom, for example, the controllability of the s-wave scattering length by the Feshbach resonance. In the case of ultracold dipolar atoms, droplet formation is known to be related to the supersolid states in which spatial translational symmetry is broken\cite{Ilzhoefer2019}. Such supersolid states have been actively studied in recent years\cite{L_onard_2017_nature,L_onard_2017_science,Morales_2018}. The recently realized supersolid state in optical lattice~\cite{Kovrizhin_2005} is well described by the Bose-Hubbard model. In the earlier study~\cite{Danshita_2009}, various ordered phases and unstable regions were found in the Bose-Hubbard system with the dipolar interaction.

	In this study, we establish a toy model describing droplets in the ``unstable'' region of the dipole Bose-Hubbard system. We use our model to estimate the size of droplets and discuss characteristic features of the shape and size of droplets. We also compare the results of the toy model with the results of numerical calculations and discuss the validity of the toy model. This research provides a qualitative understanding of droplet formation and shape control of droplets.

\section{Ordered phases of Dipolar Bose-Hubbard model}\label{sec:Ordered_phases_of_Dipolar_Bose_Hubbard_model}
	The Hamiltonian of the dipolar Bose-Hubbard model is given by
	
	\begin{align}
		\hat{\mathcal{H}}=-\sum_{\braket{jk}}J_{jk}\hat{b}^\dagger_j\hat{b}_k+\frac{1}{2}\sum_jU_j\hat{b}^\dagger_j\hat{b}_j(\hat{b}^\dagger_j\hat{b}_j-1)+\frac{1}{2}\sum_{\braket{jk}}U_{jk}\hat{b}^\dagger_j\hat{b}_j\hat{b}^\dagger_k\hat{b}_k,
	\end{align}
	where $\braket{jk}$ is nearest neighbor sites. The last term is the dipolar interaction between nearest neighbor sites.

%不要な図と二国先生に指摘されたので削除
%	\begin{wrapfigure}{r}[0\linewidth]{0.5\linewidth}
%		\centering
%		%\includegraphics[width=0.6\linewidth]{./wave_moji.eps}
%		\includegraphics[width=1.0\linewidth]{./wave_moji.eps}
%		\caption{Schematic diagram of dipolar Bose-Hubbard model}
%		\label{fig:Schematic_diagram_of_dipolar_Bose-Hubbard_model}
%	\end{wrapfigure}

	The previous study~\cite{Danshita_2009} showed the emergence of three quantum phases, SF(superfluid), SSS(striped supersolid), and CSS(checker-board supersolid), and an unstable region as a ground state of the two-dimensional dipolar Bose-Hubbard model.

	The energy of the dipolar Bose-Hubbard model in the mean-field approximation is given by
	\begin{align}
	E =-\sum_{\braket{jk}}J_{jk}\phi^*_j\phi_k+\frac{1}{2}U\sum_j\lvert\phi_j\rvert^4 + \frac{1}{2}\sum_{\braket{jk}}U_{jk}\lvert\phi_j\rvert^2\lvert\phi_k\rvert^2,
	\end{align}
	where the order parameter is defined as an expectation value of the annihilation operator $\phi_j=\braket{\hat{b}_j}$. The order parameter of the three  quantum phases can be written as
	\begin{align}
		\begin{cases}
			\phi_j^{\text{SF}}=\sqrt{n},\\
			\phi_j^{\text{SSS(x)}}=\sqrt{n}_0+\sqrt{n}_1\exp(i\bm{k}_x\cdot\bm{r}_j/2),\\
			\phi_j^{\text{CSS}}=\sqrt{n}_0+\sqrt{n}_1\exp(i\bm{k}_{xy}\cdot\bm{r}_j/2),
		\end{cases}
		\label{eq:ordered_phase_order_parameter}
	\end{align}
	where $\bm{k}_x=(1/a_x,0,0),\bm{k}_{xy}=(1/a_x,1/a_y,0)$ with the lattice constants $a_x,a_y$, and $n$ is the average number of particles.
	We can obtain a phase diagram of the dipolar Bose-Hubbard model by comparing energies of the system in the three phases with the stability condition:
	\begin{align}
		\label{eq:stability_condition_ordered}
		\frac{\partial n}{\partial \mu}>0,
	\end{align}
	where $n$ is given by $n_0+n_1$ for SSS and CSS. The chemical potential $\mu$ is defined as
	\begin{align}
		\mu = \frac{\partial E}{\partial N},
	\end{align}
	where $N$ is the total number of particles.

\begin{figure}[H]\label{fig:density_profiles}
		\begin{minipage}[b]{0.25\linewidth}
			\subcaption{\hspace*{0.8\linewidth}}
			\includegraphics[width=1\linewidth]{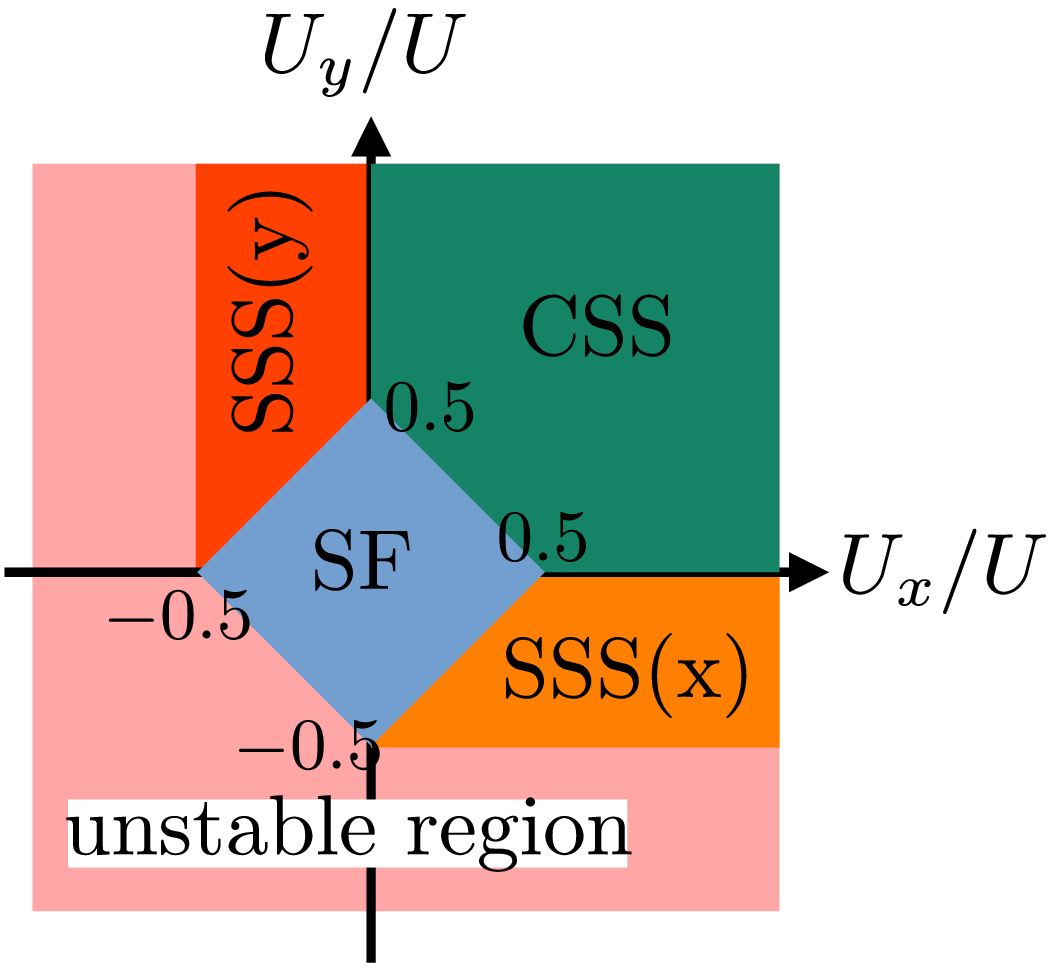}
			\label{subfig:phase_diagram}
		\end{minipage}
		\begin{minipage}[b]{0.32\linewidth}
			\subcaption{SF\hspace*{0.7\linewidth}}
			\includegraphics[width=1\linewidth]{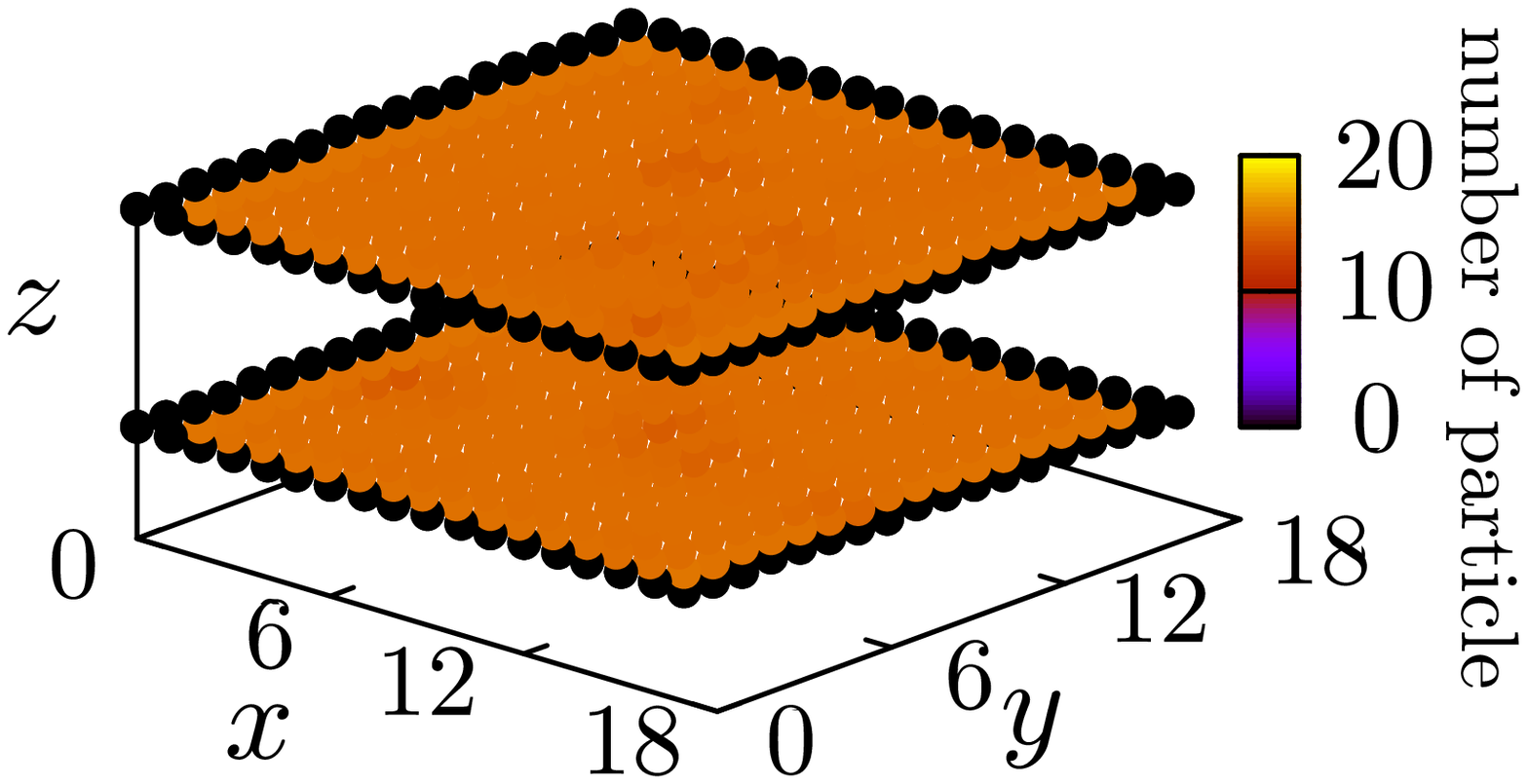}
			\label{subfig:sf_density_profile}
		\end{minipage}
		\begin{minipage}[b]{0.32\linewidth}
			\subcaption{SSS(x)\hspace*{0.5\linewidth}}
			\includegraphics[width=1\linewidth]{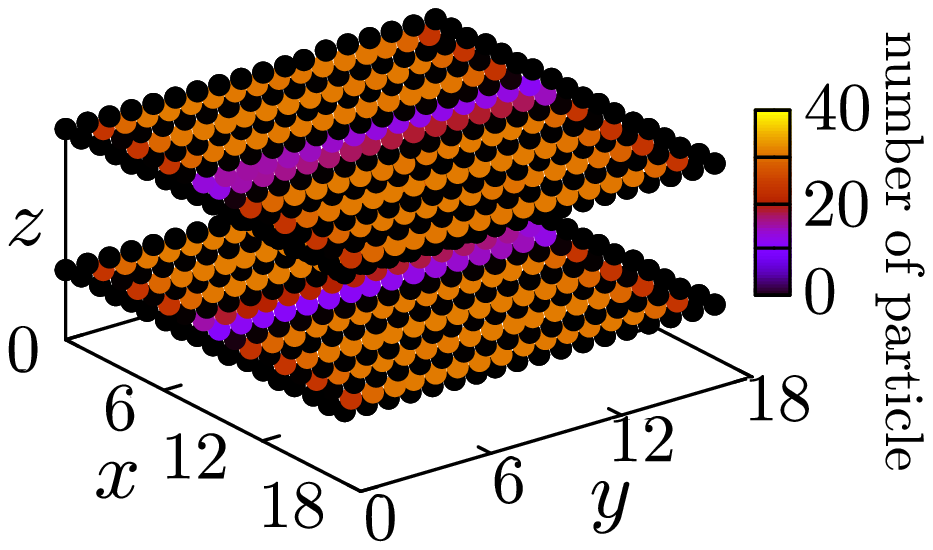}
			\label{subfig:sss(x)_density_profile}
		\end{minipage}
		\begin{minipage}[b]{0.32\linewidth}
			\subcaption{SSS(y)\hspace*{0.5\linewidth}}
			\includegraphics[width=1\linewidth]{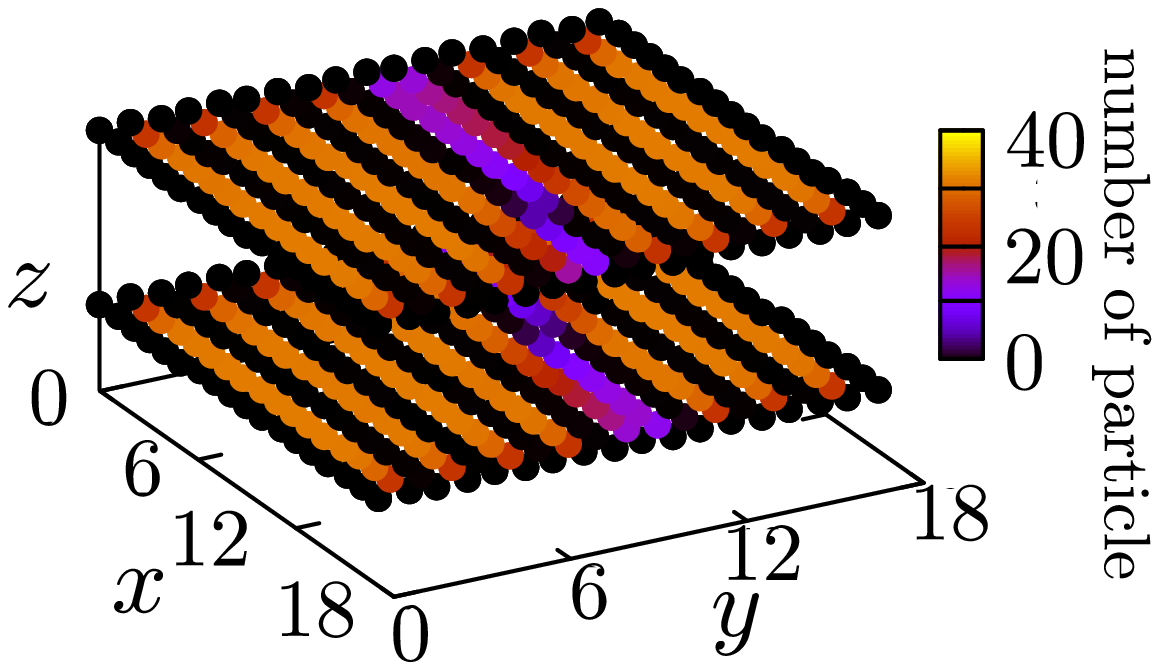}
			\label{subfig:sss(y)_density_profile}
		\end{minipage}
		\begin{minipage}[b]{0.32\linewidth}
			\subcaption{CSS\hspace*{0.6\linewidth}}
			\includegraphics[width=1\linewidth]{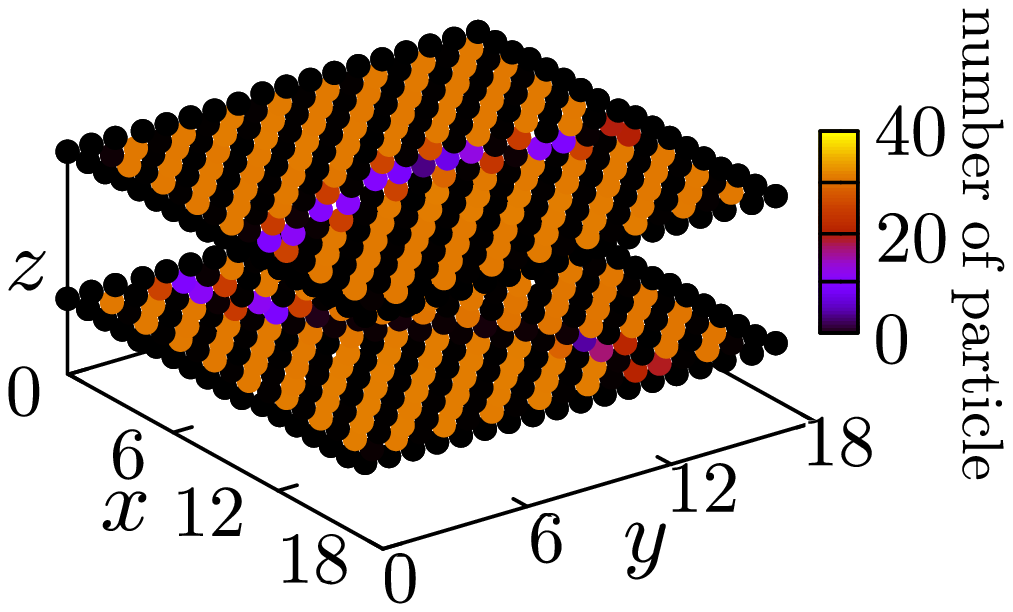}
			\label{subfig:css_density_profile}
		\end{minipage}
		\begin{minipage}[b]{0.32\linewidth}
			\subcaption{Unstable region\hspace*{0.2\linewidth}}
			\includegraphics[width=1\linewidth]{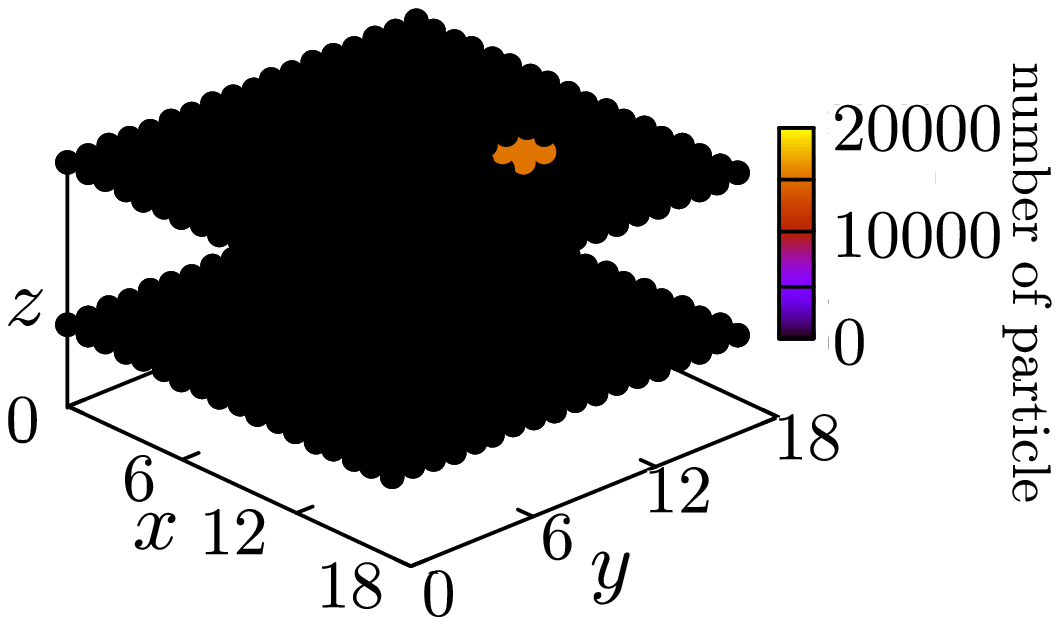}
			\label{subfig:unstable_density_profile}
		\end{minipage}
		\caption{Density profiles of ordered phases of dipolar Bose-Hubbard model. The hopping parameter is set as $J_x=0.1U,J_y=0.2U,J_z=0.05U$ and the interaction strength is set as $U_z=-0.05U$. The density and size of the system are set as $N/V=15$ and $V=16\times16\times16$. (\subref{subfig:phase_diagram}) Analytically obtained phase diagram of ordered phases: (\subref{subfig:sf_density_profile}) SF region $(U_{x,y}=0.025U)$, (\subref{subfig:sss(x)_density_profile}) SSS(x) region $[(U_x,U_y)=(0.525,-0.225)U]$, (\subref{subfig:sss(y)_density_profile}) SSS(y) region $[(U_x,U_y)=(-0.225,0.525)U]$, (\subref{subfig:css_density_profile}) CSS region $[(U_x,U_y)=(0.575U,0.575U)]$ and (\subref{subfig:unstable_density_profile}) Unstable region $(U_{x,y}=-1.275U)$.}
	\end{figure}

	Figure~\ref{subfig:phase_diagram} shows the phase diagram.
%	\begin{wrapfigure}{r}[0pt]{0.5\linewidth}
%		\centering
%		\includegraphics[width=1.0\linewidth]{./previous_research_phase_diagram.eps}
%		\caption[phase diagram of dipolar Bose-Hubbard model in 2 dimension]{phase diagram of dipolar Bose-Hubbard model in 2 dimension}
%		\label{fig:phase_diagram_of_dipolar_Bose-Hubbard_model_in_2_dimension}
%	\end{wrapfigure}
	The phase boundary of the phase diagram can be determined by comparing the energy and analyzing the stability condition~(\ref{eq:stability_condition_ordered}).
	Previous study~\cite{Danshita_2009} has shown that the particles are uniformly distributed when the dipolar interaction is weak; the SSS state is achieved when the dipolar interaction in one direction is stronger in the other direction; the CSS state is achieved when the dipolar interactions in both directions are repulsive and strong (Fig.~\ref{subfig:phase_diagram}).

	We also determine the ground state of the system numerically using the imaginary time evolution method without assuming the order parameter in the form of Eq.~(\ref{eq:ordered_phase_order_parameter}). Figures~\ref{subfig:sf_density_profile}-\ref{subfig:unstable_density_profile} show the ground state obtained by numerical calculations. The initial distribution of the imaginary time evolution method is set as a random distribution with an average of 15 particles at each site. Open boundary conditions are applied.

	The SF state has a spatially uniform distribution; the SSS state has a spatially periodic distribution in one direction; the CSS state has a spatially periodic distribution in two directions.
	Actually, there are some inhomogeneities because of the open boundary condition. The inhomogeneity is less significant in CSS because, in this state, the occupied lattice site is surrounded by vacant lattice sites in the $xy$ plane. In contrast, in SSS the inhomogeneity appears along the stripe.
	The numerical calculations with imaginary-time evolution show that droplet states emerge in the so-called ``unstable region''. These droplets are thin in the $z$ direction and spread over a few sites in the $x$-$y$ plane. In later sections, we establish a toy model of these droplet states and compare them with the numerical results.

\section{Box-droplet model in dipolar Bose-Hubbard model}\label{sec:Box-droplet_model_in_Dipolar_Bose-Hubbard_model}

The droplets can emerge in the ``unstable region'' of the dipolar Bose-Hubbard model. Here, we assume the ``box-droplet'' with a uniform density distribution inside the droplet and no particles outside. The order parameter of the box droplet can be written as

	\begin{align}\label{eq:box_droplet_order_parameter}
		\phi_j=\begin{cases}
			\displaystyle\sqrt{\frac{N}{l_xl_yl_z}}&\displaystyle\left( \frac{\|x_j\|}{\sigma_x},\frac{\|y_j\|}{\sigma_y},\frac{\|z_j\|}{\sigma_z}\le 1 \right)\\
			0&\text{otherwise,}
		\end{cases},
	\end{align}
	where $x_j,y_j,z_j$ are site indices and $N$ is the total number of particles. We assumed that the center of the droplet is located at the origin of the system $x_j=y_j=z_j=0$. $\sigma_x,\sigma_y,\sigma_z$ are the ``radius'' of the droplet and $l_x,l_y,l_z$ are the ``thickness'' of the droplet, which is defined as $l_i=2\sigma_i+1$ for $i=x,y,z$. The droplet should have at least a single site thickness in each direction, and thus $1\le l_{x,y,z}$.

	The mean-field energy of the box-droplet is given by
	\begin{align}
		\label{eq:energy_of_droplet}
		E_{\mathrm{BOX}}=&-2\left( J_x\frac{l_x-1}{l_x}+ J_y\frac{l_y-1}{l_y}+J_z\frac{l_z-1}{l_z}\right)N\nonumber\\
		&\ +\frac{N^2}{2l_xl_yl_z}\left( U+2U_x\frac{l_x-1}{l_x} +2U_y\frac{l_y-1}{l_y}+2U_z\frac{l_z-1}{l_z}\right).
	\end{align}
	If the system size is much greater than the size of the droplet, one can neglect the first term of Eq.(\ref{eq:energy_of_droplet}) as
	\begin{align}
		E_{\mathrm{BOX}}\simeq&\frac{N^2}{2l_xl_yl_z}\left( U+2U_x\frac{l_x-1}{l_x} +2U_y\frac{l_y-1}{l_y}+2U_z\frac{l_z-1}{l_z}\right).
		\label{eq:energy_of_droplet_without_first_term}
	\end{align}
	The width of the droplet is obtained by minimizing the energy~(\ref{eq:energy_of_droplet_without_first_term}). For $\|U_z\|\ll U$, we obtain the following simple formulas for the droplet width:

	\begin{align}
		l_x=
		\begin{dcases}
			1&\left(U_y<-U/2+2U_x\right),\\
			\frac{6U_x}{U+2(U_x+U_y)}&\left(-U/2+2U_x<U_y<\frac{2U_x+U}{4}\right),\\
			\frac{4U_x}{U+2U_x}&\left(\frac{2U_x+U}{4}<U_y\right).
		\end{dcases}
		\label{eq:droplet_width_of_box_droplet}
	\end{align}
	The analogous expression for $l_y$ can be obtained by exchanging $x$ and $y$ in Eq.(\ref{eq:droplet_width_of_box_droplet}).

\begin{figure}[H]\label{fig:droplets}
		\begin{minipage}[b]{0.32\linewidth}
			\subcaption{\hspace*{0.8\linewidth}}
			\includegraphics[width=1\linewidth]{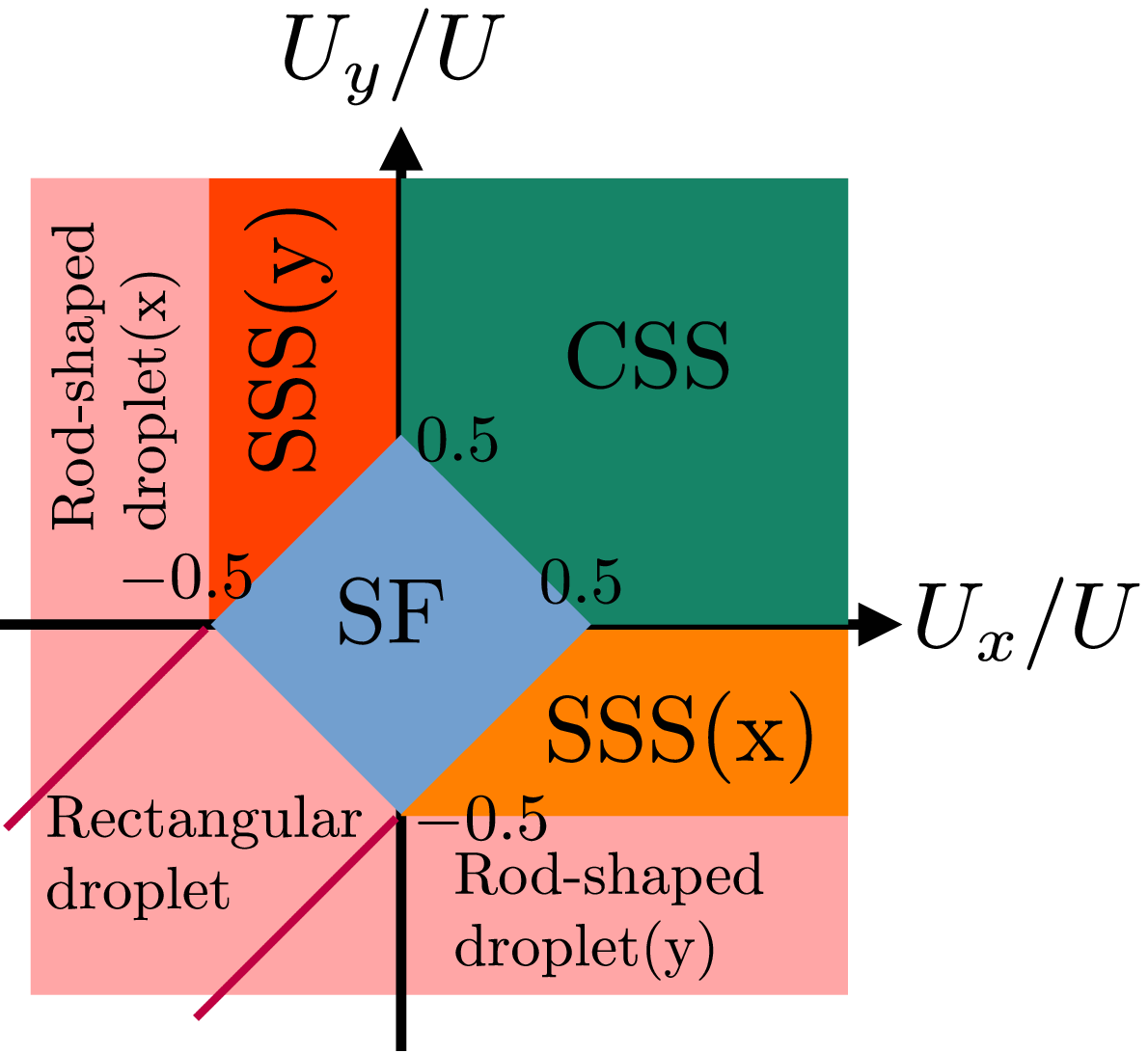}
			\label{subfig:droplet_phase_diagram}
		\end{minipage}
		\begin{minipage}[b]{0.32\linewidth}
			\subcaption{Rectangular droplet}
			\includegraphics[width=1\linewidth]{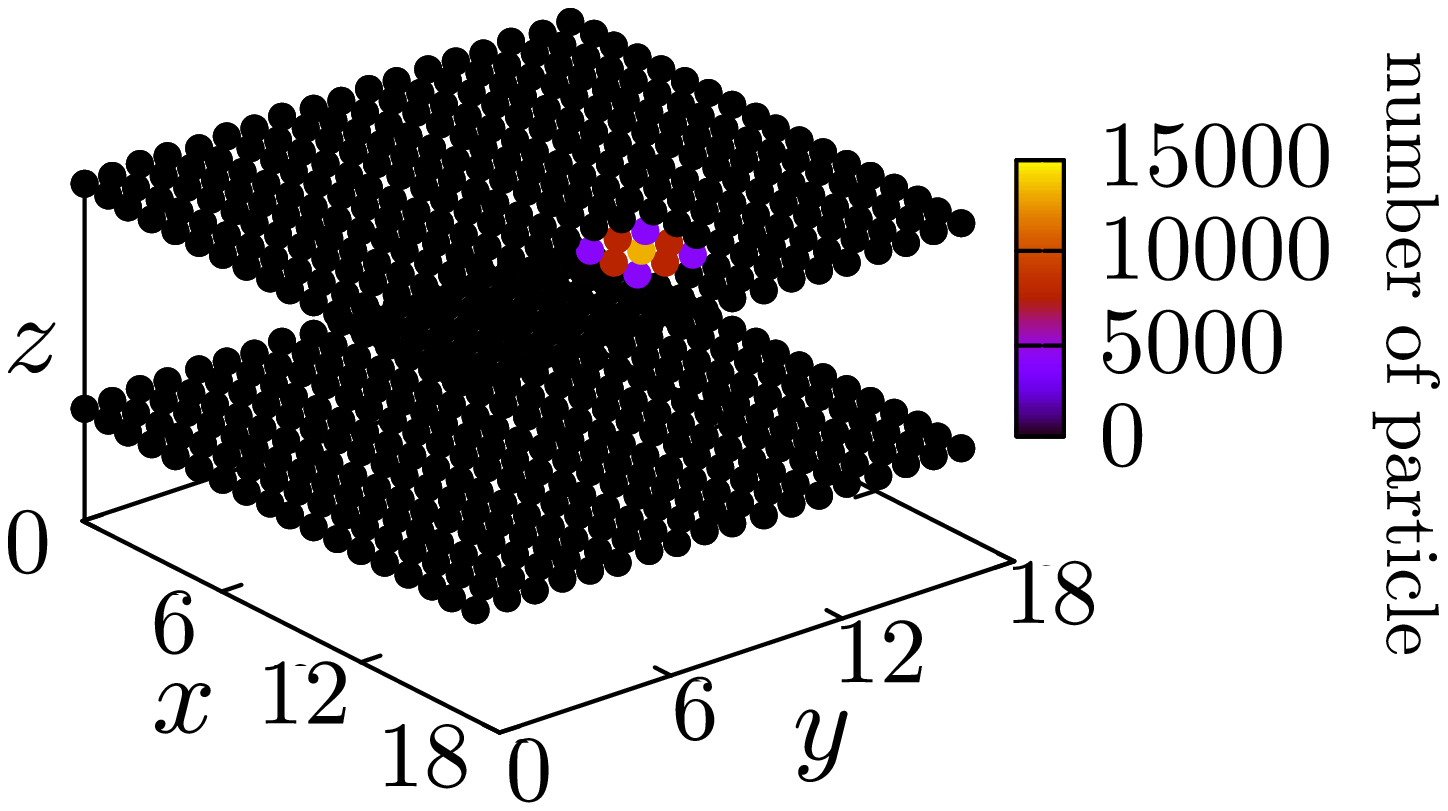}
			\label{subfig:rectangular_droplet_density_profile}
		\end{minipage}
		\begin{minipage}[b]{0.32\linewidth}
			\subcaption{Rod-shaped droplet}
			\includegraphics[width=1\linewidth]{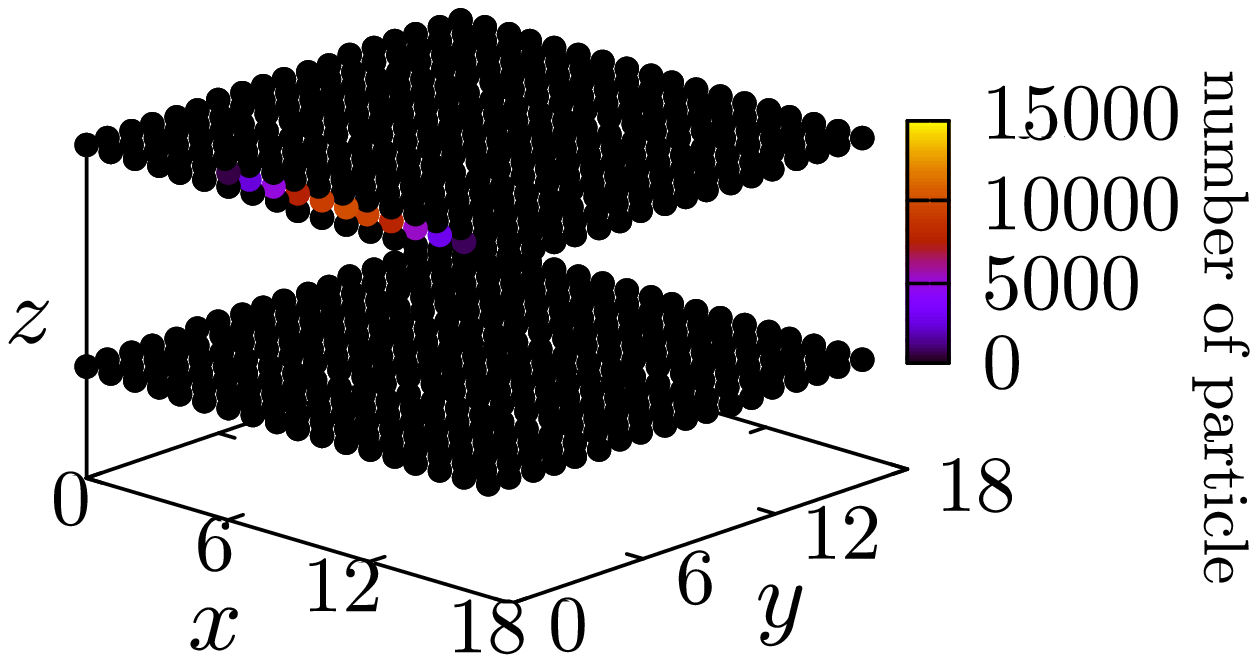}
			\label{subfig:rod_shaped_droplet(x)_density_profile}
		\end{minipage}
		\caption{Density profiles of droplets of dipolar Bose-Hubbard model. The hopping parameters are set as $J_x=0.1U,J_y=0.2U,J_z=0.05U$ and the interaction strength is set as $U_z=-0.05U$. The density and size of the system are set as $N/V=15$ and $V=16\times16\times16$. (\subref{subfig:droplet_phase_diagram}) Phase diagram with droplet phase, (\subref{subfig:rectangular_droplet_density_profile}) Rectangular droplet with $U_{x,y}=-0.575U$ and (\subref{subfig:rod_shaped_droplet(x)_density_profile}) Rod-shaped droplet with $(U_x,U_y)=(-0.575,0.575)U$.}
	\end{figure}

	The droplets described by this toy model can be roughly classified into two types: rod-shaped droplet and rectangular droplet.
	The rod-shaped droplet oriented in the $x$ direction has a single site thickness in the $y$ and $z$ directions and a width of more than one site in the $x$ direction. On the other hand, the rectangular droplet has a single site thickness in the $z$ direction and has a width of more than one site in the $x$ and $y$ directions.
	Comparing the energy of the box-droplet with that of the ordered phases, we find that the droplet state has lower energy in the unstable region that satisfies $\partial n / \partial \mu < 0$. Figure~\ref{subfig:droplet_phase_diagram} shows the phase diagram of the dipolar Bose-Hubbard model including the droplet phase obtained by using the box-droplet model. We also show the density profiles of droplet phases obtained by the numerical calculation in Figs.~\ref{subfig:rectangular_droplet_density_profile} and \ref{subfig:rod_shaped_droplet(x)_density_profile}.

\section{Droplet width of box-droplet model}

\begin{figure}[H]
		{\subcaption*{\leftline{\hspace*{2em}(a)Box-droplet\hspace{0.25\linewidth}(b)Numerical}}}
		\includegraphics[width=0.9\linewidth]{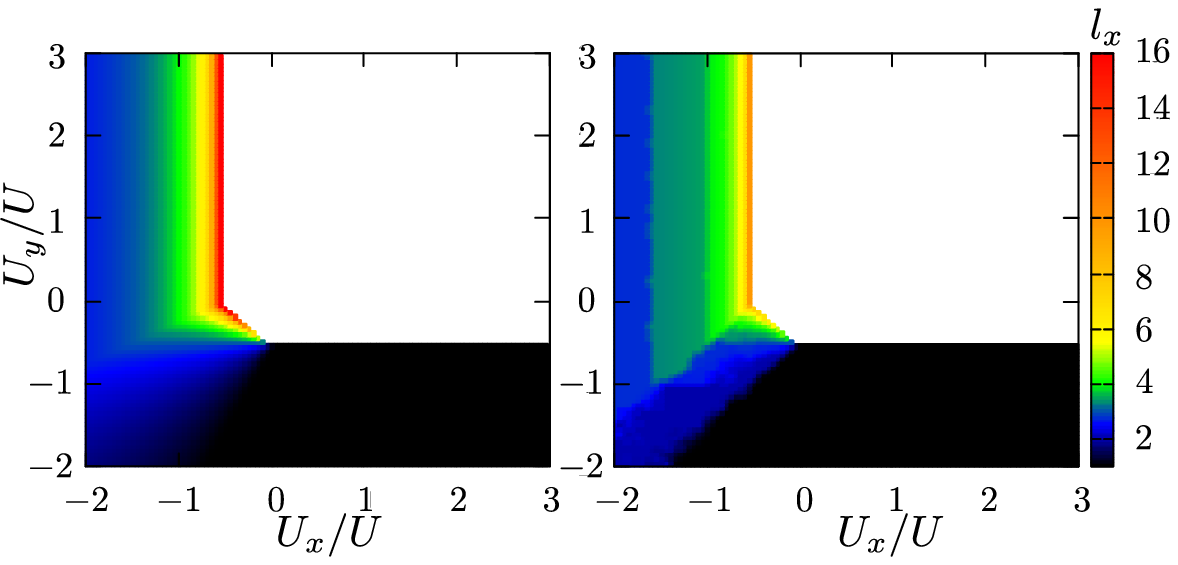}\\
		\caption{Droplet width of the box-droplet model and numerical calculation. The hopping parameter and interaction strength are set as $J_x=0.1U,J_y=0.2U,J_z=0.05U$, and $U_z=-0.05U$. Droplet width of (a) the box-droplet model and (b) numerical calculation.}
		\label{fig:droplet_width_comparison}
	\end{figure}

Figure~\ref{fig:droplet_width_comparison} compares the width of the box-droplet with that of the numerical calculation. The droplet width obtained by the box-droplet model is plotted in Fig.~\ref{fig:droplet_width_comparison}(a) and the numerical result is shown plotted in Fig.~\ref{fig:droplet_width_comparison}(b).
We also show the droplet width estimated from the numerical results and that of our box-droplet model in Table\ref{tab:comparison_of_droplet_width}.

	\begin{table}[h]
		\centering
		\begin{tabular}{|ll|cc|cc|c|}
			\hline
			& & \multicolumn{2}{|c|}{Box-droplet} & \multicolumn{2}{|c|}{Numerical} & Droplet type\\
			$U_x/U$ & $U_y/U$ & $l_x$ & $l_y$ & $l_x$ & $l_y$ &\\
			\hline\hline
			-1.975&1.025&2.678&1.000&2.748&1.000&Rod-shaped droplet(x)\\
			-0.975&1.025&4.105&1.000&3.908&1.000&Rod-shaped droplet(x)\\
			-0.575&1.025&15.33&1.000&7.666&1.000&Rod-shaped droplet(x)\\
			-1.975&-1.975&2.000&2.000&1.717&1.717&Rectangular droplet\\
			-0.975&-0.975&2.017&2.017&2.037&2.639&Rectangular droplet\\
			-0.575&-0.575&2.654&2.654&2.715&2.715&Rectangular droplet\\
			-0.275&-0.275&16.50&16.50&7.367&7.354&Rectangular droplet\\
%			以下は古いデータ
%			-1.25&1.25&4.33&1.00&3.37&1.00\\
%			-0.55&-0.55&2.75&2.75&2.75&2.75\\
%			-0.55&0.55&22.0&1.00&9.11&1.00\\
%			-0.55&1.25&22.0&1.00&8.84&1.00\\
%			 0.35&-0.35&5.25&5.25&3.99&4.12\\
			\hline
		\end{tabular}
		\caption{droplet width}
		\label{tab:comparison_of_droplet_width}
	\end{table}

	In most parameter regions, the box-droplet model can accurately predict the numerically calculated droplet width. Only in the narrow region near the boundary between ordered and droplet phases, the numerically calculated width of the droplet becomes very large and the box-droplet model becomes less accurate.
	According to the numerical calculation, the Box-droplet model describes the shape of droplets reasonably well when the droplet width is smaller than 5. This condition corresponds to $U_x\le-0.83U$ for the rod-shaped droplet oriented in the $x$ direction and $U_y\le-0.4U_x-0.5U$ for the rectangular droplet.
	This difference between the numerical calculation and the box-droplet model can be understood from the density distribution of the droplet.
	The distribution is approximately uniform if the droplet is small enough. In contrast, the density distribution cannot be considered uniform if the droplet width is much greater than one site.

\section{Conclusion}
We have calculated the ground state of the dipolar Bose-Hubbard system by using the imaginary time evolution method. As a result, we found that the droplet exists in the ``unstable region'', which is consistent with the previous study. We constructed a box-droplet model that can qualitatively predict the shape of droplets. The box-droplet model is less accurate for large droplets where the density distribution cannot be considered uniform. In contrast, the box-droplet model can accurately predict the droplet width for small droplets.

\section*{Acknowledgment}
This work is supported by JSPS KAKENHI Grants No. JP18K03499.


\begin{thebibliography}{1}
\providecommand{\url}[1]{{#1}}
\providecommand{\urlprefix}{URL }
\providecommand{\doi}[1]{\url{https://doi.org/#1}}
\bibcommenthead

\bibitem{Ilzhoefer2019}
P.~Ilzh^^c3^^b6fer, M.~Sohmen, G.~Durastante, C.~Politi, A.~Trautmann,
  G.~Natale, G.~Morpurgo, T.~Giamarchi, L.~Chomaz, M.J. Mark, and F.~Ferlaino,
  Nat. Phys. \textbf{17}, 356 (2021).

\bibitem{L_onard_2017_nature}
J.~L{\'{e}}onard, A.~Morales, P.~Zupancic, T.~Esslinger, and T.~Donner, Nature
  \textbf{543}, 87 (2017).

\bibitem{L_onard_2017_science}
J.~L{\'{e}}onard, A.~Morales, P.~Zupancic, T.~Donner, and T.~Esslinger, Science
  \textbf{358}, 1415 (2017).

\bibitem{Morales_2018}
A.~Morales, P.~Zupancic, J.~L{\'{e}}onard, T.~Esslinger, and T.~Donner, Nat.
  Mater. \textbf{17}, 686 (2018).

\bibitem{Kovrizhin_2005}
D.L. Kovrizhin, G.V. Pai, and S.~Sinha, Europhys. Lett. \textbf{72}, 162
  (2005).

\bibitem{Danshita_2009}
I.~Danshita, and C.A.R. {{\relax S{\'a}} de Melo}, Phys. Rev. Lett.
  \textbf{103}, 225301 (2009).

\end{thebibliography}
\end{document}